\documentstyle[12pt,epsfig,graphics,cite]{article}
\begin{document}\bibliographystyle{plain}\begin{titlepage}
\renewcommand{\thefootnote}{\fnsymbol{footnote}}\hfill\begin{tabular}{l}
HEPHY-PUB 800/05\\May 2005\end{tabular}\\[1.5cm]\Large
\begin{center}{\bf RELATIVISTIC HARMONIC OSCILLATOR}\\\vspace{1.5cm}
\large{\bf LI Zhi-Feng\footnote[3]{\normalsize\ {\em Present address\/}:
Institute for Theoretical Physics, University of Vienna, Boltzmanngasse 5,
A-1090 Vienna, Austria}}\\[.2cm]\normalsize Department of Physics, Chongqing
University, 400044 Chongqing, China\\[.5cm]
\large{\bf LIU Jin-Jin}\\[.2cm]\normalsize Department of Modern Physics,
University of Science and Technology of China, Hefei, China\\[.5cm]
\large{\bf Wolfgang LUCHA\footnote[1]{\normalsize\ {\em E-mail address\/}:
wolfgang.lucha@oeaw.ac.at}}\\[.2cm]\normalsize Institute for High Energy
Physics, Austrian Academy of Sciences,\\Nikolsdorfergasse 18, A-1050 Vienna,
Austria\\[.5cm]\large{\bf MA Wen-Gan}\\[.2cm]\normalsize Department of Modern
Physics, University of Science and Technology of China, Hefei, China\\[.5cm]
\large{\bf Franz F.~SCH\"OBERL\footnote[2]{\normalsize\ {\em E-mail
address\/}: franz.schoeberl@univie.ac.at}}\\[.2cm]\normalsize Institute for
Theoretical Physics, University of Vienna,\\Boltzmanngasse 5, A-1090 Vienna,
Austria\vfill{\normalsize\bf Abstract}\end{center}\normalsize We study the
semirelativistic Hamiltonian operator composed of the relativistic kinetic
energy and a static harmonic-oscillator potential in three spatial dimensions
and construct, for bound states with vanishing orbital angular momentum, its
eigenfunctions in ``compact form,'' i.e.,~as power series, with expansion
coefficients determined by an explicitly given recurrence relation. The
corresponding eigenvalues are fixed by the requirement of normalizability of
the solutions.\vspace{.5cm} {\em PACS numbers\/}: 03.65.Pm, 03.65.Ge
\renewcommand{\thefootnote}{\arabic{footnote}}\end{titlepage}

\normalsize\section{Introduction: \hspace{-1pt}relativistic
\hspace{-1pt}harmonic-oscillator \hspace{-1pt}problem}The simplest and
perhaps most straightforward generalization of the Schr\"odinger operators of
standard nonrelativistic quantum theory towards the inclusion of relativistic
kinematics leads to Hamiltonians $H$ that involve the relativistic kinetic
energy, or relativistically covariant form of the free energy, of a particle
of mass $m$ and momentum $\mbox{\boldmath{$p$}},$ given by the square-root
operator$$T(p)\equiv\sqrt{p^2+m^2}\ ,\qquad p\equiv|\mbox{\boldmath{$p$}}|\
,$$and a coordinate-dependent static interaction potential
$V(\mbox{\boldmath{$x$}}).$ In the one-body case, they~read\begin{equation}
H=\sqrt{p^2+m^2}+V(\mbox{\boldmath{$x$}})\ .\label{Eq:SRH}\end{equation}The
eigenvalue equation of this Hamiltonian is usually called the ``spinless
Salpeter equation.'' It may be regarded as a well-defined approximation to
the Bethe--Salpeter formalism \cite{BSE}~for~the description of bound states
within relativistic quantum field theories, obtained when assuming that all
bound-state constituents interact {\em instantaneously\/} and propagate like
{\em free\/} particles \cite{SE}. Among others, it yields semirelativistic
descriptions of hadrons as bound states of quarks
\cite{Lucha91:BSQ,Lucha92:QAQBS}.

In general, the above semirelativistic Hamiltonian $H$ is, unfortunately, a
{\em nonlocal\/} operator: either the relativistic kinetic energy, $T(p),$
{\em in configuration space\/} or, in general, the interaction potential {\em
in momentum space\/} is a nonlocal operator. Because of the nonlocality it is
somewhat difficult to obtain rigorous {\em analytical\/} statements about the
solutions of its eigenvalue equation. Thus sophisticated methods have been
developed to extract information about these solutions; for details and
comparisons of the various approaches, consult, for example, the reviews
\cite{Lucha94:Como,Lucha98:Dubna,Lucha:Oberwoelz,Lucha:Dubrovnik,Lucha04:TWR,
Lucha04:CP}.

Analytical or, at least, semianalytical\footnote{We regard a bound on some
eigenvalue of a given self-adjoint operator as {\em semianalytical\/} if it
can be~derived by the --- in general, numerical --- optimization of an
analytically known expression over a single real variable.} expressions for
both upper and lower bounds on the eigenvalues of some self-adjoint operator
may be found by combining the minimum--maximum principle
\cite{Reed78,Weinstein72,Thirring90} and appropriate operator inequalities
\cite{Martin88,Martin89,Lucha96:rcpaubel,Lucha99:1dimsrcp,Lucha99:A/Q}. The
outcome of this procedure is sometimes called the ``spectral comparison
theorem.'' In Sec.~\ref{Sec:EE} below, we will use this kind of bounds in
order to estimate the accuracy of our findings for the eigenvalues of the
operator (\ref{Eq:SRH}). Accordingly, we recall in Appendix~\ref{App:SCT} the
proof of the ``translation'' of some inequality satisfied by two operators
into the corresponding relations between their discrete eigenvalues, by
briefly sketching all basic assumptions and the line of argument. A very
systematic path for obtaining such operator inequalities is provided by
rather simple geometrical considerations summarized under the notion
``envelope theory'' \cite{Lucha00:HO,Lucha01:DMAIa,Lucha01:DMAIb,Lucha02:sum,
Lucha02:DMAII,Lucha02:CP}. These envelope techniques may be generalized to
systems composed of arbitrary numbers of relativistically moving interacting
particles \cite{Lucha01:NHO,Lucha03:NHO-m=0,Lucha04:NV(r2)}. For particular
potentials $V,$ semianalytical lower bounds on the ground-state energy
eigenvalue of the semirelativistic Hamiltonian (\ref{Eq:SRH}), and hence on
the entire spectrum of $H,$ can be found by the appropriate generalization of
the local-energy theorem \cite{Duffin47,Barnsley78,Baumgartner79,Thirring90}
to our case of relativistic kinematics \cite{Raynal94,Lucha02:DMAII}, or by
applying the optimized (Beckner--Brascamp--Lieb) version of Young's
inequality for convolutions to some integral formulation of the spinless
Salpeter equation \cite{Brau04}.

Purely numerical solutions of the spinless Salpeter equation may be computed
in numerous ways. The semirelativistic Hamiltonian $H$ can be approximated by
some effective Hamiltonian which is of nonrelativistic shape but uses
parameters that depend on expectation values of the momenta
\cite{Lucha94:ESRH,Lucha95:ESRH}. Upper bounds of, in principle, arbitrarily
high precision on the eigenvalues of a self-adjoint operator can be found
\cite{Jacobs86,Lucha94:VA-SRCP,Lucha96:CCC,Lucha97:LagB,Lucha99:Q/A,
Lucha99:A/Q} with the help of the Rayleigh--Ritz (variational) technique as
immediate consequence of the minimum--maximum theorem
\cite{Reed78,Weinstein72,Thirring90}. The spinless Salpeter equation may also
be converted into an equivalent matrix eigenvalue problem
\cite{Nickisch84,Gara90,Lucha92:MAP,Fulcher93,Fulcher94,Lucha00:SAMM}.

A particular r\^ole for $H$ is played by the spherically symmetric
harmonic-oscillator potential$$V(\mbox{\boldmath{$x$}})=a\,r^2\ ,\qquad
r\equiv|\mbox{\boldmath{$x$}}|\ ,\qquad a>0\ ;$$this potential defines the
``relativistic harmonic-oscillator problem,'' posed by the Hamiltonian
\begin{equation}H=\sqrt{p^2+m^2}+a\,r^2\ .\label{Eq:RHOP}\end{equation}The
eigenvalue equation of $H,$ for eigenstates $|\psi\rangle,$
$H\,|\psi\rangle=E\,|\psi\rangle,$ involves only one parameter: upon
factorizing off some overall energy scale $a^{1/3}$ by performing the
canonical transformation$$r\to\frac{r}{a^{1/3}}\ ,\qquad p\to a^{1/3}\,p$$and
rescaling both mass $m$ and eigenvalue $E$ according to $m=a^{1/3}\,\mu$ and
$E=a^{1/3}\,\varepsilon,$ it~reads
$$\left(\sqrt{p^2+\mu^2}+r^2\right)|\psi\rangle=\varepsilon\,|\psi\rangle\
.$$In momentum-space representation this eigenvalue equation reduces to a
Schr\"odinger
problem\begin{equation}[-\Delta_p+V(p)]\,\psi(p)=\varepsilon\,\psi(p)\
,\label{Eq:MSR}\end{equation}with an interaction potential reminiscent of the
square root of the relativistic kinetic energy:$$V(p)\equiv\sqrt{p^2+\mu^2}\
.$$

We would like to take advantage of this facet of the relativistic
harmonic-oscillator problem in order to derive for its bound-state
eigenfunctions, in this analysis, a compact expression and to move thereby
beyond only numerically calculated exact solutions without having to rely on
perturbation theory, maybe paving the way for the eventual construction of
analytic~solutions.

\section{Analytical solutions for bound-state eigenfunctions}\label{Sec:SAA}
We exploit the fact that, in contrast to the general case, for a
harmonic-oscillator potential (as, with due care, for any potential of the
form $V\propto r^{2n},$ $n=1,2,3,\dots$) the eigenvalue~equation of the
Hamiltonian $H$ is an ordinary differential equation, parametrized by $\mu$
and its eigenvalue~$\varepsilon.$

Focusing on the ground state or purely radial excitations, let us introduce,
for eigenstates with vanishing relative orbital angular momentum $\ell,$ the
reduced radial wave function $y(p)$~by$$y(p)\equiv\sqrt{4\,\pi}\,p\,\psi(p)
\qquad(\ell=0)\ ,$$which is, of course, subject to the normalization
condition$$\int\limits_0^\infty{\rm d}p\,|y(p)|^2=1\ .$$In accordance with
Eq.~(\ref{Eq:MSR}) the reduced radial wave function satisfies a reduced
radial equation:\begin{equation}\frac{{\rm d}^2y}{{\rm d}p^2}(p)=
[V(p)-\varepsilon]\,y(p)\ .\label{Eq:RSE}\end{equation}From its definition,
this {\em reduced\/} radial wave function $y(p)$ has to vanish at the origin:
$y(0)=0.$ Moreover, the analysis of the {\em normalizable\/} solutions of the
eigenvalue equation (\ref{Eq:RSE}) reveals~that $y(p)$ behaves like $p$ for
small $p,$ that is, for $p\ll1$; hence, its derivative with respect to $p$ at
the point $p=0$ is a {\em nonvanishing\/} constant, which may be absorbed
into the overall normalization:$$\frac{{\rm d}y}{{\rm d}p}(0)=1\ .$$

We construct all solutions of Eq.~(\ref{Eq:RSE}) in form of Taylor-series
expansions by using the~ansatz$$y(p)=\sum_{n=0}^\infty
c_n\,\frac{p^n}{n!}$$with the expansion coefficients$$c_n\equiv\frac{{\rm
d}^ny}{{\rm d}p^n}(0)\ ,\qquad n=0,1,2,\dots\ .$$The solution of the
eigenvalue equation (\ref{Eq:RSE}) is then clearly tantamount to the
determination~of the expansion coefficients $c_n.$ The first three of these
expansion coefficients are known trivially:\begin{eqnarray*}c_0&=&y(0)=0\
,\\[1ex]c_1&=&\frac{{\rm d}y}{{\rm d}p}(0)=1\ ,\\[1ex]c_2&=&\frac{{\rm
d}^2y}{{\rm d}p^2}(0)=[(V-\varepsilon)\,y](0)=0\ .\end{eqnarray*}[For the
sake of notational simplicity, we suppress, in accordance with our above
remark, in the following that normalization factor of $y(p)$ which guarantees
its unity norm and assume~$y(p)$ to be normalized such that the value of the
first nonvanishing expansion coefficient is one: $c_1=1.$] Upon insertion of
the eigenvalue equation (\ref{Eq:RSE}) followed by the application of
Leibniz's theorem, the nontrivial expansion coefficients $c_n,$ $n\ge 3,$ may
be shown to satisfy the recurrence
relation:\begin{eqnarray}c_{n+2}&=&\frac{{\rm d}^{n+2}y}{{\rm
d}p^{n+2}}(0)=\frac{{\rm d}^n}{{\rm d}p^n}\left[\frac{{\rm d}^2y}{{\rm
d}p^2}\right](0)=\frac{{\rm d}^n[(V-\varepsilon)\,y]}{{\rm
d}p^n}(0)\nonumber\\[1ex]&=&\sum_{k=0}^n
\left(\begin{array}{c}n\\k\end{array}\right)\left[\frac{{\rm
d}^k(V-\varepsilon)}{{\rm d}p^k}\,\frac{{\rm d}^{n-k}y}{{\rm
d}p^{n-k}}\right](0)\nonumber\\[1ex]&=&(\mu-\varepsilon)\,c_n+
\sum_{k=1}^n\left(\begin{array}{c}n\\k\end{array}\right)
\frac{{\rm d}^kV}{{\rm d}p^k}(0)\,c_{n-k}\nonumber\\[1ex]
&=&(\mu-\varepsilon)\,c_n+
\sum_{k=1}^n\left(\begin{array}{c}n\\k\end{array}\right)\mu^{1-k}\,d_k\,c_{n-k}
\ ,\qquad n=1,2,3,\dots\ .\label{Eq:RR1}\end{eqnarray}Here, for the last
step, we abbreviated the $k$-th derivative of the potential $V$ by a
coefficient~$d_k$:$$d_k\equiv\mu^{k-1}\,\frac{{\rm d}^kV}{{\rm
d}p^k}(0)=\frac{{\rm d}^k\sqrt{x^2+1}}{{\rm d}x^k}(0)\ ,\qquad
x\equiv\frac{p}{\mu}\ ,\qquad\mu>0\ ,\qquad k=0,1,2,\dots\ .$$In the case
$\mu=m=0,$ the solutions involve Airy's function ${\rm Ai}(z);$ cf., e.~g.,
Refs.~\cite{Lucha94:ESRH,Lucha95:ESRH,Lucha99:A/Q,Lucha02:DMAII}. According
to the above definition, we have $d_0=1.$ Furthermore, by inspection of the
function $f(x)=\sqrt{x^2+1},$ it is easy to convince oneself that all odd
derivatives of $f(x)$ vanish at
$x=0,$\begin{equation}d_{2k+1}=0\qquad\mbox{for all}\ k=0,1,2,\dots\
,\label{Eq:do0}\end{equation}whereas all even derivatives of $f(x)$ at $x=0$
necessarily satisfy the (simple) recurrence~relation
$$d_{2k+2}=(1-4\,k^2)\,d_{2k}\ ,\qquad k=0,1,2,\dots\ .$$By induction, the
solution of this recurrence relation for the nonvanishing coefficients
$d_{2k}$~reads\begin{equation}d_{2k}=(-1)^{k-1}\,(2\,k-1)
\left[\frac{(2\,k-2)!}{2^{k-1}\,(k-1)!}\right]^2\ ,\qquad k=1,2,3,\dots\
.\label{Eq:DS}\end{equation}Taking into account the observation
(\ref{Eq:do0}), we obtain $c_3=\mu-\varepsilon$ and, for the coefficients
$c_n,$~$n\ge4,$$$c_{n+2}=(\mu-\varepsilon)\,c_n+\sum_{k=1}^{[n/2]}
\left(\begin{array}{c}n\\2\,k\end{array}\right)\mu^{1-2k}\,d_{2k}\,c_{n-2k}\
,\qquad n=2,3,4,\dots\
,$$where$$\left[\frac{n}{2}\right]\equiv\left\{\begin{tabular}{ll}
$\displaystyle\frac{n}{2}$&$\quad\mbox{for $n$ even, $n=2,4,6,\dots$}\
,$\\[2ex]$\displaystyle\frac{n-1}{2}$&$\quad\mbox{for $n$ odd,
$n=3,5,7,\dots$}\ .$\end{tabular}\right.$$Thus the recurrence relation
(\ref{Eq:RR1}) for all expansion coefficients $c_n$ decomposes into one
involving only the even coefficients $c_{2n},$ $n=2,3,4,\dots,$ and one
involving only the odd coefficients $c_{2n+1},$ $n=2,3,4,\dots;$ recalling
$c_0=0$ and $c_2=0,$ we conclude that all the {\em even\/} coefficients $c_n$
vanish:$$c_{2n}=0\qquad\mbox{for all}\ n=0,1,2,\dots\ .$$With the result
(\ref{Eq:DS}), the recurrence relation for the (nonvanishing) {\em odd\/}
coefficients finally~reads
\begin{eqnarray}c_{2n+3}&=&(\mu-\varepsilon)\,c_{2n+1}+\sum_{k=1}^n
\left(\begin{array}{c}2\,n+1\\2\,k\end{array}\right)\mu^{1-2k}\,d_{2k}\,
c_{2n-2k+1}\nonumber\\[1ex]&=&(\mu-\varepsilon)\,c_{2n+1}+\sum_{k=1}^n
\left(\begin{array}{c}2\,n+1\\2\,k\end{array}\right)c_{2n-2k+1}\,\mu^{1-2k}\,
(-1)^{k-1}\,(2\,k-1)\left[\frac{(2\,k-2)!}{2^{k-1}\,(k-1)!}\right]^2\
,\nonumber\\[1ex]&&\qquad n=1,2,3,\dots\ .\label{Eq:RR2}\end{eqnarray}In
summary, upon constructing the relevant expansion coefficients according to
this recurrence relation the analytical expressions for all reduced radial
wave functions (of $\ell=0$ bound states) $y(p)$ of the relativistic
harmonic-oscillator problem (\ref{Eq:RHOP}) are given by the power-series
expansion\begin{equation}y(p)=\sum_{n=0}^\infty
c_{2n+1}\,\frac{p^{2n+1}}{(2\,n+1)!}\ .\label{Eq:RWF}\end{equation}

The various solutions $y(p)$ of the eigenvalue equation (\ref{Eq:MSR}) are
characterized or discriminated by different sets of expansion coefficients
$c_n.$ By construction, apart from the first coefficient $c_1,$ all expansion
coefficients for a given solution depend on the corresponding energy
eigenvalue $\varepsilon.$ Evaluating the recurrence relation (\ref{Eq:RR2})
for just the first term, the series (\ref{Eq:RWF}) explicitly starts~with
$$y(p)=p\left\{1+(\mu-\varepsilon)\,\frac{p^2}{3!}+
\left[(\mu-\varepsilon)^2+\frac{3}{\mu}\right]\frac{p^4}{5!}+\dots\right\}.$$

\section{Explicit bound-state eigenfunctions and energy levels}\label{Sec:EE}
The central result of our present investigation of the ``relativistic
harmonic-oscillator problem'' defined by the Hamiltonian $H$ of
Eq.~(\ref{Eq:RHOP}) is an analytical expression for the reduced radial~parts
$y(p)$ of all eigenfunctions $\psi(p)$ of $H$ for vanishing angular momentum
in form of a Taylor series:$$y(p)\equiv\sqrt{4\,\pi}\,p\,\psi(p)
=\sum_{n=0}^\infty c_{2n+1}\,\frac{p^{2n+1}}{(2\,n+1)!}\ ,$$where the
corresponding expansion coefficients $c_{2n+1},$ $n=0,1,2,\dots,$ are either
fixed to~$c_1=1,$ $c_3=\mu-\varepsilon$ or, for $c_{2n+1},$ $n\ge2,$
determined by the recurrence relation (\ref{Eq:RR2}). For a given~solution of
this eigenvalue problem these expansion coefficients and thus the resulting
reduced radial~wave function $y(p)$ depend on the parameter $\mu\equiv
m/a^{1/3}$ and the corresponding energy eigenvalue $\varepsilon.$ Taking into
account the necessary requirement of normalizability of Hilbert-space
eigenstates, the inversion of the latter relation may be exploited to
determine the energy eigenvalues $\varepsilon$ from the knowledge of the
dependence of $\varepsilon$ on the coefficients $c_n,$ as derived in the
preceding~section. In order to fulfil its normalization condition, any
solution $y(p)$ must vanish in the limit $p\to\infty$:
\begin{equation}\lim_{p\to\infty}y(p)=0\ .\label{Eq:NCC}\end{equation}Fixing
the energy eigenvalue $\varepsilon$ of a chosen bound state in this manner
and using this particular value of the parameter $\varepsilon$ in the
expansion (\ref{Eq:RWF}) then yields the corresponding wave function~$y(p).$

Our principal concern is beyond doubt the semianalytical approach developed
in Sec.~\ref{Sec:SAA}~and summarized above. Nevertheless, it might be
instructive to construct explicitly a few examples of solutions in numerical
or graphical form. These results can be compared with the outcome of some
straightforward (but merely numerical) integration of the Schr\"odinger
equation (\ref{Eq:RSE}).~This will provide a useful check of the correctness
of our solutions and justify the present formalism.

In actual computations, the infinite series (\ref{Eq:RWF}) has to be
truncated, for practical purposes, to a reasonably large but definitely
finite number $N$ of terms considered in this expansion~of~$y(p)$:
\begin{equation}y(p)\simeq\sum_{n=0}^N c_{2n+1}\,\frac{p^{2n+1}}{(2\,n+1)!}\
.\label{Eq:PST}\end{equation}In this case, the wave function $y(p)$ will
approach zero, as required by the constraint (\ref{Eq:NCC}),~not at infinity
but already for a finite value, say $\widehat p,$ of the momentum $p$ ---
before it starts to diverge. This technique gives the energy eigenvalues
$\varepsilon$ with a precision determined by one's choice~of~$N$:
$\varepsilon=\varepsilon(N).$ Likewise, the momentum boundary $\widehat p$
will also change with $N$: $\widehat p=\widehat p(N).$ For a given value of
$\mu,$ which quantifies the relative importance of particle mass $m$ and
harmonic-oscillator coupling strength $a,$ every wave function resulting from
this truncation procedure involves~two dimensionless parameters: the relevant
energy eigenvalue $\varepsilon(N)$ of the Hamiltonian (\ref{Eq:RHOP}), and
the characteristic momentum $\widehat p(N).$ Their values will be determined
simultaneously, in accordance with the above requirement on $y(p)$ [to
approach zero at $\widehat p(N)$] by an appropriate fit procedure. In other
words, the value of $\widehat p(N),$ in particular, cannot be varied freely;
it is fixed for chosen~$N.$

\begin{figure}[ht]\begin{center}\psfig{figure=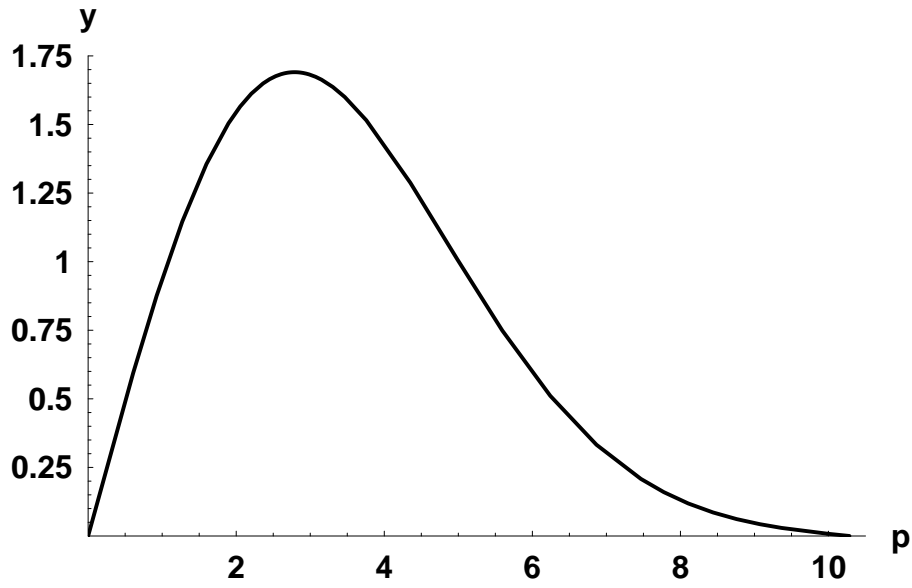,scale=1.194}\\[1ex](a)
\\[1ex]\psfig{figure=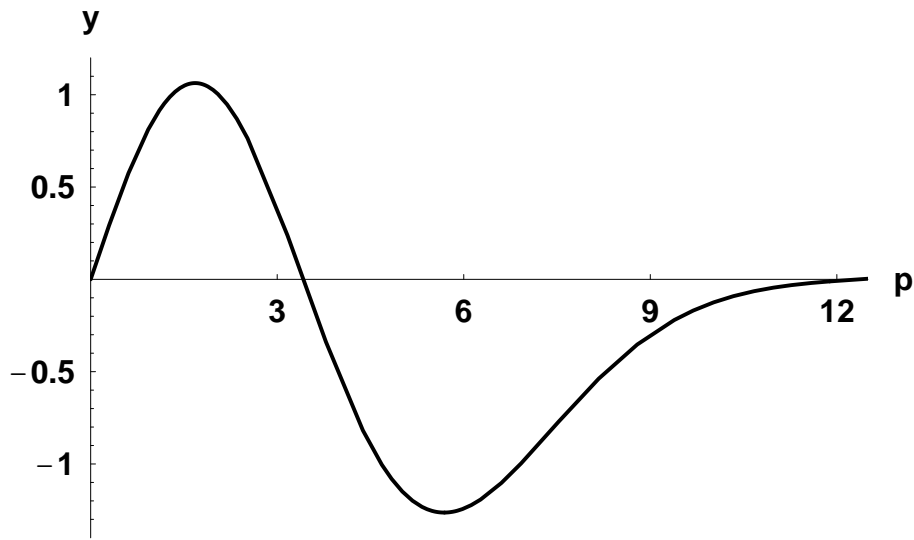,scale=1.2}\\[1ex](b)\caption{Radial
eigenfunctions in momentum space $y(p)\equiv\sqrt{4\,\pi}\,p\,\psi(p)$ of the
ground state~(a) and the first radial excitation (b) of the relativistic
harmonic-oscillator problem, defined by~the Hamiltonian
$H=\sqrt{p^2+\mu^2}+r^2,$ with $\mu=30,$ and for 45 terms in the Taylor
expansion of~$y(p).$}\label{Fig:RHO44}\end{center}\end{figure}

Let us illustrate this procedure for both ground state and first radial
excitation, i.~e., for the two bound states defined by vanishing orbital
angular momentum and radial quantum~number $n_{\rm r}=0,1,$ resp.
Figure~\ref{Fig:RHO44} shows the corresponding reduced radial wave functions
$y(p)$ for $\mu=30$ as obtained by inspecting the functional form of $y(p)$
resulting from different choices~of~$\varepsilon$~and~$\widehat p,$ if taking
into account 45 terms in their Taylor series (\ref{Eq:RWF}), that is, if
choosing $N=44$ in~Eq.~(\ref{Eq:PST}). Table~\ref{Tab:ES} summarizes the
relevant numerical parameter values emerging from such construction.

\begin{table}[ht]\caption{Dimensionless (by scaling) energy eigenvalues
$\varepsilon(N)$ and characteristic momenta $\widehat p(N)$ of the
relativistic harmonic-oscillator problem posed by the Hamiltonian
$H=\sqrt{p^2+\mu^2}+r^2$ as well as the classical turning points $\overline
p$ of the corresponding nonrelativistic motion for the ground state and the
first radial excitation (identified by their radial quantum number $n_{\rm
r}=0,1$),~with mass-vs.-coupling strength parameter $\mu=30,$ and $N=44$ or
$N=14$ in our Taylor series~(\ref{Eq:PST}).}\label{Tab:ES}
\begin{center}\begin{tabular}{ccccc}\hline\hline&\\[-1.5ex]
\multicolumn{1}{c}{$N$}&\multicolumn{1}{c}{$n_{\rm r}$}&
\multicolumn{1}{c}{$\varepsilon(N)-\mu$}&
\multicolumn{1}{c}{$\widehat p(N)$}&\multicolumn{1}{c}{$\overline p$}\\[1ex]
\hline\\[-1.5ex]44&0&0.38627&10&4.82\\&1&0.89864&12&7.36\\[1ex]\hline\\[-1.5ex]
14&0&0.38854&8.5&4.82\\[1ex]\hline\hline\end{tabular}\end{center}\end{table}

The exact results can be easily computed with the aid of a (standard)
integration technique designed for solving the Schr\"odinger equation
numerically \cite{Lucha98:Num}. For our two examples the exact wave functions
$y(p)$ prove to be practically indistinguishable, at least by the eye,
from~the ones extracted from the series (\ref{Eq:PST}) with $N=44.$ This
explains why we refrain from plotting also the former in
Fig.~\ref{Fig:RHO44}. For the momentum range depicted in
Fig.~\ref{Fig:RHO44}, that is, for $0\le p\le\widehat p,$~the~relative
differences of the areas under corresponding curves are (of the order)
$10^{-8};$ more precisely,~they are given by $2\times10^{-8}$ for $n_{\rm
r}=0,$ the ground state, and $3\times10^{-8}$ for $n_{\rm r}=1,$ the first
excited~level.

From a straightforward consideration of the ``classical turning points'' of
the corresponding (and well understood) nonrelativistic harmonic-oscillator
problem defined by the Hamiltonian$$H_{\rm NR}=\mu+\frac{p^2}{2\,\mu}+r^2\
,$$the maximum ``classical'' momenta, $\overline p,$ are found, in terms of
the radial quantum number~$n_{\rm r},$~as$$\overline p^2=(4\,n_{\rm
r}+3)\,\sqrt{2\,\mu}\ ,\quad n_{\rm r}=0,1,2,\dots\ .$$By inspecting
Table~\ref{Tab:ES} we note with satisfaction that for both energy levels
under consideration the numerical values of the suitable $\widehat p$ turn
out to be far beyond their classical counterparts~$\overline p.$

\begin{figure}[ht]\begin{center}\psfig{figure=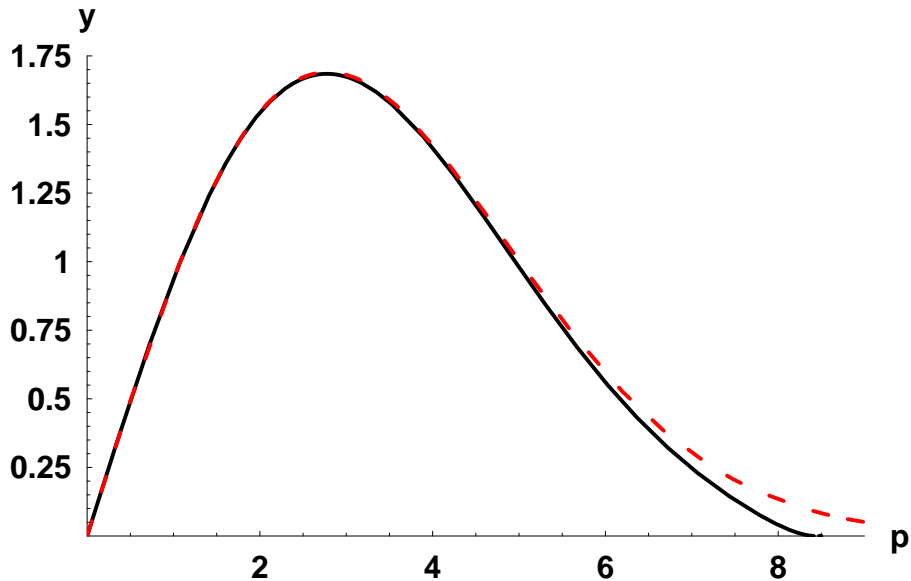,scale=1.194}\\[1ex]
\caption{Momentum-space wave function $y(p)$ of Fig.~\ref{Fig:RHO44}(a)
resulting from consideration of only 15 terms in its polynomial approximation
(\ref{Eq:PST}) [{\sl full line\/}], in comparison with the corresponding
exact ground-state wave function of the relativistic harmonic-oscillator
operator [{\sl dashed line\/}].}\label{Fig:RHO14}\end{center}\end{figure}

In order to get, at least, some vague idea of the dependence of our findings
on the amount of truncation represented by $N<\infty,$ we inspect the
ground-state wave function $y(p)$ constructed again for $\mu=30$ but by
truncating the expansion (\ref{Eq:RWF}) to the rather modest number of 15
terms, which means to set $N=14$ in Eq.~(\ref{Eq:PST}).
Figure~\ref{Fig:RHO14} confronts this approximate wave function $y(p)$ with
its exact behaviour for the ground state. While for small and
intermediate~momenta there is still perfect agreement with the exact result
\cite{Lucha98:Num}, we observe a clearly discernible~discrepancy between
approximate and exact curve for large momenta. Table~\ref{Tab:ES} tells us
that even for~$N=14$ our crucial momentum $\widehat p$ is still comfortably
above the corresponding classical turning point, $\overline p.$ Moreover,
comparing the cases $N=14$ and $N=44,$ we learn that the value of $\widehat
p$ increases~with increasing number $N.$ Of course, the naive expectation
would be that $\widehat p$ behaves like $\widehat p\to\infty$ for
$N\to\infty,$ that is, when removing the truncation and restoring the full
series expansion~for~$y(p).$

The minimum number of terms to be taken into account in the Taylor-series
expansions (\ref{Eq:RWF}) required in order to achieve some given precision
of one's results will depend, of course,~on~both the bound state under study
and the desired accuracy. From our above remarks we feel entitled to conclude
that a reasonable (and manageable) number $N\simeq40$ produces satisfactory
results.

To our knowledge, at present the best {\em semianalytical\/} upper and lower
bounds to the energy eigenvalues of the relativistic harmonic-oscillator
problem are provided simultaneously by the combination of minimum--maximum
principle with operator inequalities \cite{Lucha99:A/Q} and the envelope
theory \cite{Lucha00:HO,Lucha01:DMAIb}: at least for the relativistic
harmonic oscillator the envelope bounds \cite{Lucha00:HO,Lucha01:DMAIb} may
be shown \cite{Lucha02:DMAII} to be quantitatively equivalent to the bounds
derived in Appendix~A of Ref.~\cite{Lucha99:A/Q}. However, a discussion, in
full generality, of all implications of such operator inequalities~for the
eigenvalues of the operators considered appears clearly off the mainstream of
our presentation. Therefore, as already promised in the Introduction, the
general relationship is demonstrated in Appendix~\ref{App:SCT}. The bounds we
need here are derived from this general theorem by specializing to the case
of the relativistic harmonic-oscillator problem posed by the Hamiltonian
operator (\ref{Eq:RHOP}); all operator inequalities required by this
procedure may be generated by, e.~g., envelope~theory. For a bound state of
vanishing relative orbital angular momentum $\ell,$ that is, for a purely
radial excitation, identified by the radial quantum number $n_{\rm
r}=0,1,2,\dots$ (identical to the number~of nodes of the corresponding wave
function), the bounds on the dimensionless eigenvalue $\varepsilon$
read$$\min_{r>0}\left(\sqrt{\mu^2+\frac{P_{\rm L}^2}{r^2}}+r^2\right)\le
\varepsilon\le\min_{r>0}\left(\sqrt{\mu^2+\frac{P_{\rm
U}^2}{r^2}}+r^2\right),$$where, in three spatial dimensions, our
envelope-theory upper-bound parameter $P_{\rm U}$ is given~by$$P_{\rm
U}=2\,n_{\rm r}+\frac{3}{2}\ ,\quad n_{\rm r}=0,1,2,\dots\ ,$$while the
lower-bound parameter $P_{\rm L}$ required for our envelope bounds is related
to the zeros $z_0$ of Airy's function ${\rm Ai}(z)$ \cite{Abramowitz}
($-z_0=2.338107,\,4.087949,\,5.520560,\,6.786708,\,7.944134,\,\dots$)~by
$$P_{\rm L}=2\left(\frac{-z_0}{3}\right)^{3/2}\ ,\quad{\rm Ai}(z_0)=0\ .$$
Resulting values of $P_{\rm L}$ for the lowest-lying $\ell=0$ bound states
are listed in Table~\ref{Tab:PL}
\cite{Lucha00:HO,Lucha01:DMAIb,Lucha02:sum,Lucha02:DMAII,Lucha04:TWR}.

\begin{table}[ht]\caption{Numerical values of the parameter $P_{\rm L}$ used
by envelope theory for the lower bounds on the energy levels of the
relativistic harmonic-oscillator problem in three spatial dimensions, for the
lowest-lying $\ell=0$ bound states identified by their radial quantum number
$n_{\rm r}=0,1,2,\dots.$}\label{Tab:PL}
\begin{center}\begin{tabular}{cr}\hline\hline&\\[-1.5ex]
\multicolumn{1}{c}{$n_{\rm r}$}&\multicolumn{1}{c}{$P_{\rm L}$}\\[1ex]
\hline\\[-1.5ex]0&1.3760835\\1&3.1813129\\2&4.9925543\\3&6.8051369\\
4&8.6182269\\[1ex]\hline\hline\end{tabular}\end{center}\end{table}

Table~\ref{Tab:ERHO} compares, for the lowest four $\ell=0$ energy levels
(identified by their radial quantum number $n_{\rm r}=0,1,2,3$) of our
Hamiltonian (\ref{Eq:RHOP}), the approximate eigenvalues, $\varepsilon(N),$
obtained by the present approach by a truncation of the power series
(\ref{Eq:PST}) to $N=14$ or $N=44$ terms,~with the corresponding
semianalytical upper ($\varepsilon_{\rm U}$) and lower ($\varepsilon_{\rm
L}$) bounds mentioned above as well as with the (numerically exact)
eigenvalues $\varepsilon_{\rm num},$ computed by a method developed for
the~purely numerical solution of (nonrelativistic) Schr\"odinger equations
\cite{Lucha98:Num}. For $N=14,$ the polynomial in $\varepsilon$ resulting
from the suitably adapted boundary condition (\ref{Eq:NCC}) has only two real
roots at all. Moreover, both of these approximate values are still above our
(semianalytical) upper~bounds. In contrast to this rather crude
approximation, for $N=44$ the eigenvalues $\varepsilon(N)$ of already the
three lowest energy levels fit perfectly to the ranges spanned by the
semianalytical bounds. For the ground state, in particular, $\varepsilon(44)$
reproduces the exact result at least to five decimal~places.

\begin{table}[ht]\caption{``Compact-origin'' eigenvalues $\varepsilon(N),$
their upper ($\varepsilon_{\rm U}$) and lower ($\varepsilon_{\rm L}$) bounds,
and~their exact values $\varepsilon_{\rm num}$ for the lowest $\ell=0$ states
of the Hamiltonian $H/a^{1/3}$ in Eq.~(\ref{Eq:RHOP}), with $\mu=30.$
}\label{Tab:ERHO}\begin{center}\begin{tabular}{lcccc}\hline\hline&&\\[-1.5ex]
\multicolumn{1}{l}{Radial Excitation $n_{\rm r}$}&\multicolumn{1}{c}{0}&
\multicolumn{1}{c}{1}&\multicolumn{1}{c}{2}&\multicolumn{1}{c}{3}
\\[1ex]\hline\\[-1.5ex] $\varepsilon_{\rm
U}-\mu$&0.38668&0.90032&1.41179&1.92111\\ $\varepsilon_{\rm
L}-\mu$&0.35478&0.81862&1.28222&1.74440\\ $\varepsilon_{\rm
num}-\mu$&0.38627&0.89857&1.40768&1.91366\\
$\varepsilon(N=14)-\mu$&0.38854&0.93616&---&---\\
$\varepsilon(N=44)-\mu$&0.38627&0.89864&1.41032&1.94319\\[1ex]
\hline\hline\end{tabular}\end{center}\end{table}

\section{Summary and Conclusions}Our compact result for the reduced $\ell=0$
eigenfunctions of the Hamiltonian (\ref{Eq:RHOP}) is given by the power
series (\ref{Eq:RWF}), with expansion coefficients $c_1=1,$
$c_3=\mu-\varepsilon,$ and $c_{2n+1},$ $n\ge2,$ determined by the recurrence
relation (\ref{Eq:RR2}). Both the numerical determination of all energy
eigenvalues and the explicit construction of the corresponding eigenfunctions
of the relativistic harmonic-oscillator problem is then achieved by forcing
our solutions to satisfy, in addition, the constraint imposed by the
requirement of normalizability of bound-state wave functions. Comparing these
explicit solutions with the outcomes of purely numerical integration
procedures reveals that at least for the lowest-lying energy levels our
semianalytical approach reproduces, already for a truncation of the Taylor
series (\ref{Eq:RWF}) to a moderate number of expansion terms, the exact
solutions with high accuracy. Of course, if one is interested only in
numerical solutions of the problem~under~study, their straightforward
computation with the help of some integration algorithm should produce the
desired result more easily than their extraction from our Taylor series by
means of Eq.~(\ref{Eq:NCC}).

\section*{Acknowledgements}One of us (F.~F.~S.) would like to thank the
University of Science and Technology of China~of the Chinese Academy of
Sciences, and the Department of Physics of the Chongqing University for
hospitality during his stay in China, during which part of the present work
has been done.

\appendix\section{Combination of minimum--maximum principle with operator
inequality: ``spectral comparison theorem''}\label{App:SCT}It is a simple
exercise to relate discrete eigenvalues of two operators satisfying some
inequality.

Consider for some operator $A,$ with domain ${\cal D}(A),$ its eigenvalue
equation $A\,|\alpha_k\rangle=a_k\,|\alpha_k\rangle,$ $k=0,1,2,\dots,$ for
its set of eigenstates $\{|\alpha_k\rangle,\ k=0,1,2,\dots\},$ corresponding
to its eigenvalues$$a_k\equiv\frac{\langle\alpha_k|\,A\,|\alpha_k\rangle}
{\langle\alpha_k|\alpha_k\rangle}\ ,\qquad k=0,1,2,\dots\ ,$$and likewise for
some operator $B,$ with domain ${\cal D}(B),$ its eigenvalue equation
$B\,|\beta_k\rangle=b_k\,|\beta_k\rangle,$ $k=0,1,2,\dots,$ for its set of
eigenstates $\{|\beta_k\rangle,\ k=0,1,2,\dots\},$ corresponding to its
eigenvalues$$b_k\equiv\frac{\langle\beta_k|\,B\,|\beta_k\rangle}
{\langle\beta_k|\beta_k\rangle}\ ,\qquad k=0,1,2,\dots\ .$$Assume that both
these operators $A$ and $B$ are self-adjoint: $A^\dagger=A$, $B^\dagger=B.$
This implies~that all their eigenvalues are real: $a_k^\ast=a_k,$
$b_k^\ast=b_k,$ $k=0,1,2,\dots.$ Let these eigenvalues be~ordered according
to $a_0\le a_1\le a_2\le\cdots,$ $b_0\le b_1\le b_2\le\cdots.$ Consider only
the discrete eigenvalues $a_k$ of $A$ below the onset of the essential
spectrum of the operator $A.$ Assume that the operator~$A$~is bounded from
below. Assume that the operators $A$ and $B$ are related by an operator
inequality of the form $A\le B,$ which implies that $B$ too is bounded from
below. In order to derive, for any $k=0,1,2,\dots,$ the relationship between
$a_k$ and $b_k,$ focus on some arbitrary $(k+1)$-dimensional subspace
$D_{k+1}$ of the domain ${\cal D}(A)$ of $A$: $D_{k+1}\subseteq{\cal D}(A).$
Employing the appropriate form of~the minimum--maximum principle, the
operator inequality $A\le B$ translates into an upper bound on the eigenvalue
$a_k$ of $A$ which involves all expectation values of $B$ within this
subspace $D_{k+1}$:\begin{equation}a_k\le\sup_{|\psi\rangle\in D_{k+1}}
\frac{\langle\psi|\,A\,|\psi\rangle}{\langle\psi|\psi\rangle}\le
\sup_{|\psi\rangle\in D_{k+1}}\frac{\langle\psi|\,B\,|\psi\rangle}
{\langle\psi|\psi\rangle}\qquad\mbox{for all}\ k=0,1,2,\dots\
.\label{Eq:MMP+OI}\end{equation}

Now, in order to relate the supremum over the expectation values of $B$ to
the eigenvalues~$b_k$ of $B,$ consider a particular subspace $D_{k+1},$
namely, that space that is spanned by the first $k+1$ eigenvectors of the
operator $B,$ that is, by precisely those eigenvectors of $B$ that correspond
to the first $k+1$ eigenvalues $b_0,b_1,\dots,b_k$ of $B$:
$D_{k+1}\subseteq{\cal D}(B)\subseteq{\cal D}(A).$ Then, clearly, every
$|\psi\rangle$ in $D_{k+1}$ is a linear combination of the eigenstates
$\{|\beta_i\rangle,\ i=0,1,\dots,k\}$ of $B,$ with coefficients~$c_i$:
$$|\psi\rangle=\sum_{i=0}^k c_i\,|\beta_i\rangle\qquad\mbox{for all}\
|\psi\rangle\in D_{k+1}\ .$$For any subspace $D_{k+1},$ $k=0,1,2,\dots,$ use
of this expansion of $|\psi\rangle$ yields for its norm
squared$$\langle\psi|\psi\rangle=\sum_{i=0}^k|c_i|^2\,
\langle\beta_i|\beta_i\rangle\qquad\mbox{for all}\ |\psi\rangle\in
D_{k+1}$$and, with this and $b_i\le b_k$ for all $i=0,1,\dots,k,$ an upper
bound on all expectation values~of~$B$: $$\langle\psi|\,B\,|\psi\rangle
=\sum_{i=0}^k|c_i|^2\,b_i\,\langle\beta_i|\beta_i\rangle\le
b_k\,\sum_{i=0}^k|c_i|^2\,\langle\beta_i|\beta_i\rangle=
b_k\,\langle\psi|\psi\rangle\qquad\mbox{for all}\ |\psi\rangle\in D_{k+1}\
,$$which means
\begin{eqnarray*}\frac{\langle\psi|\,B\,|\psi\rangle}{\langle\psi|\psi\rangle}
\le b_k&\quad&\mbox{for all}\ |\psi\rangle\in D_{k+1}\ ,\\[1ex]
\frac{\langle\psi|\,B\,|\psi\rangle}{\langle\psi|\psi\rangle}=b_k&\quad&
\mbox{for}\ |\psi\rangle=|\beta_k\rangle\in D_{k+1}\ .\end{eqnarray*}
Therefore the supremum of all expectation values of $B$ over $D_{k+1}$ is
just the eigenvalue $b_k$~of~$B$:$$\sup_{|\psi\rangle\in D_{k+1}}
\frac{\langle\psi|\,B\,|\psi\rangle}{\langle\psi|\psi\rangle}=b_k\qquad
\mbox{for all}\ k=0,1,2,\dots\ .$$Thus, inserting this identity in the chain
of inequalities (\ref{Eq:MMP+OI}) proves that corresponding discrete
eigenvalues $a_k,$ $b_k$ of semibounded self-adjoint operators $A,$ $B$ that
satisfy $A\le B$ are related~by$$a_k\le b_k\qquad\mbox{for all}\
k=0,1,2,\dots\ .$$

\end{document}